\pdfoutput=1
\RequirePackage{ifpdf}
\ifpdf % We are running pdfTeX in pdf mode
\documentclass[pdftex]{sigma}
\else
\documentclass{sigma}
\fi

\begin{document}

\newcommand{\arXivNumber}{1403.2080}

\allowdisplaybreaks

\renewcommand{\thefootnote}{$\star$}

\renewcommand{\PaperNumber}{074}

\FirstPageHeading

\ShortArticleName{The Soccer-Ball Problem}

\ArticleName{The Soccer-Ball Problem\footnote{This paper is a~contribution to the Special Issue on Deformations of
Space-Time and its Symmetries.
The full collection is available at \href{http://www.emis.de/journals/SIGMA/space-time.html}
{http://www.emis.de/journals/SIGMA/space-time.html}}}

\Author{Sabine HOSSENFELDER}

\AuthorNameForHeading{S.~Hossenfelder}

\Address{Nordita, KTH Royal Institute of Technology and Stockholm University,\\
Roslagstullsbacken 23, SE-106 91 Stockholm, Sweden}
\Email{\href{mailto:hossi@nordita.org}{hossil@nordita.org}}
\URLaddress{\url{http://backreaction.blogspot.com}}

\ArticleDates{Received March 11, 2014, in f\/inal form July 03, 2014; Published online July 09, 2014}

\Abstract{The idea that Lorentz-symmetry in momentum space could be modif\/ied but still remain observer-independent has
received quite some attention in the recent years.
This modif\/ied Lorentz-symmetry, which has been argued to arise in Loop Quantum Gravity, is being used as a phenomenological
model to test possibly observable ef\/fects of quantum gravity.
The most pressing problem in these models is the treatment of multi-particle states, known as the `soccer-ball problem'.
This article brief\/ly reviews the problem and the status of existing solution attempts.}


\Classification{83A05; 83C45}

\renewcommand{\thefootnote}{\arabic{footnote}}
\setcounter{footnote}{0}

\section{Introduction}

The Lorentz-group is non-compact and boosts can result in arbitrarily large blue shifts.
Regardless of how long a~wavelength appears to you, there always exists a~restframe in relative motion where this
wavelength is just a~Planck length or even below that.
This means in return that discarding short wavelengths, or high frequencies respectively, beyond any f\/inite cutof\/f
necessarily violates Lorentz-invariance.

Since the Planck energy is often expected to act as a~natural regulator at high frequencies, this brings up the question
how it can be given an observer-independent meaning.
One way to do this is to regard it as a~Lorentz-scalar, much like the mass of the W-boson in the coupling of Fermi's
theory of beta decay.
Another way to do it is to modify the action of Lorentz-transformation on momentum space so that the Planck energy is
observer-independent in the same way the speed of light is observer-independent.
This is the idea underlying deformations of special relativity ({\sc DSR}).
First proposed
in~\cite{AmelinoCamelia:2000ge,AmelinoCamelia:2002wr,KowalskiGlikman:2001gp,Magueijo:2001cr,Magueijo:2002am}, these
modif\/ications of special relativity have received much attention both because they have observable consequences and
because of their potential to illuminate the role of the Planck scale as a~regulator.
{\sc DSR} was originally conceived as a~modif\/ication of the
Lorentz-transformations that keeps the Planck energy as a~component of a~four-vector invariant.
It has later been motivated by Loop Quantum Gravity (for references see \cite[Section~5.1]{Hossenfelder:2012jw}).

However, while it is straight-forward to construct modif\/ied Lorentz-transformations for momentum space by rescaling the
momentum~\cite{Hossenfelder:2005ed}, this procedure cannot work in position space.
The Lorentz-group is the unique group that leaves the Minkowski-metric invariant with the origin as f\/ixed point, ie the
only point that is mapped unto itself.
The deformed transformation in momentum space is a~non-linear correction to the normally linear Lorentz-transformation
and singles out the zero-vector as a~preferred point that remains invariant.
That is unproblematic in momentum space because a~vanishing momentum is indeed a~special case.
However, a~point marked zero in space-time is as good as any other point, and thus accepting the special role of the
zero in position space violates observer-independence.
The group of translations in this case is no longer a~normal subgroup of the Poincar\'e group, which means that boosting
followed by a~spatial shift does not give the same result as f\/irst shifting and then boosting.
This is problematic because the spatial shift may be passive, i.e.~be merely a~relabeling of coordinates.
One is left with the conclusion that points labeled as being distant from zero have an ambiguous transformation behavior
that prevents one from precisely localizing the point, resulting in a~non-locality that can become macroscopically large
(much larger than the Planck length).

How to formulate {\sc DSR} in position space is thus an important question and one that has so far not been
satisfactorily addressed.
It was pointed out already in~\cite{AmelinoCamelia:2002vy,Hossenfelder:2006rr, Schutzhold:2003yp}, that the early
versions of DSR violate locality at large scales.
In~\cite{Hossenfelder:2009mu,Hossenfelder:2010tm} it was shown that this is the case for any theory with an
energy-dependent speed of light that also maintains observer-independence.
Partly in response to this highly problematic non-locality, recent attempts have been to reformulate {\sc DSR} as a~theory of `relative
locality'~\cite{AmelinoCamelia:2011yi,AmelinoCamelia:2011bm,AmelinoCamelia:2010qv,Carmona:2011wc,Jacob:2010vr,Smolin:2010mx, Smolin:2010xa}.

In the approach of relative locality the momentum space is not constructed from the tensor bundle over the space-time
manifold.
Instead, one starts with phase space and aims to reconstruct space-time from it.
Locality, then, can be interpreted as being observer-dependent, and relative rather than absolute.
It remains to be seen how this approach circumvents the issue pointed out
in~\cite{Hossenfelder:2009mu,Hossenfelder:2010tm} in any other way than just making the speed of light
energy-independent and thus reproducing ordinary Special Relativity.
The big problem in the relative locality approach is that, when starting from phase space rather than space-time, it is
unclear in the end how to identify a~physically meaningful coordinate system on space-time, for the very reason spelled
out above, that the behavior of coordinates now depends on unphysical labels.
However, this is an original new idea and still young, and so it can be hoped that with more study it will result in
a~satisfactory theory of {\sc DSR} in position space in the soon future.

In the following, I want to focus on a~dif\/ferent, though of course related, problem that occurs when modifying the
action of the Lorentz-group on momentum space, that is the addition of momenta.
We will use the unit convention $c=\hbar=1$ and the Planck mass $m_{\rm Pl}$ is the inverse of the Planck length $l_{\rm
Pl} = 1/m_{\rm Pl}$.

\section{The problem}

It is presently not known how to def\/ine the sum of momenta in approaches that modify Lorentz-symmetry in momentum space
and maintain observer-independence.
In a~nutshell, the problem is that the modif\/ied Lorentz-transformation, $\widetilde\Lambda$, that acts on momenta is
nonlinear, and thus the transformation of the sum of momenta is not the same as sum of the transformations of the momenta
\begin{gather*}
\widetilde{\Lambda}(\mathbf{p}_1 + \mathbf{p}_2) \neq \widetilde{\Lambda}(\mathbf{p}_1) +
\widetilde{\Lambda}(\mathbf{p}_2).
%\label{eq1}
\end{gather*}
This then ruins observer-independence which was one of the main motivations to introduce deformed Lorentz-symmetry: If
$\widetilde \Lambda$ is the unit element of the group, then the equation is fulf\/illed, and thus the restframe in which
it is fulf\/illed singles out a~preferred frame.

The f\/irst way to address this problem that comes to mind is to note that the momenta of dif\/ferent particles are elements
of dif\/ferent subspaces of the phase-space.
One could thus in principle just def\/ine the transformation acting on the sum of momenta to be the sum of the
transformations.
However, this procedure will fail for momentum conservation in interactions, in which the terms contributing to the sum
can be dif\/ferent before and after the interaction.
One thus concludes that it is necessary to def\/ine a~new, non-linear, addition law $\oplus$ for momenta that has the
property that it remains invariant under Lorentz-transformations and that can be rightfully interpreted as a~conserved
quantity.

Note that this problem is about the sum of momenta, and not necessarily about bound states or even quantum particles.
This problem appears already in the attempt to formulate a~classical theory which obeys the modif\/ied transformation
behavior.
The question is not about interacting particles or quantum superpositions, the question is what is the total momentum of
any collection of particles.
It is apparent that this is an essential point to address for any model that should reproduce known physics.

Luckily, it is possible to construct a~Lorentz-invariant new addition law $\oplus$ without too much trouble.
To see how this works, we note that given a~modif\/ied nonlinear Lorentz-transforma\-tion,~$\widetilde \Lambda$, that acts
on elements ${\bf p}$ of momentum space, it is always possible to construct a~vector ${\bf k}$ that is a~function of
${\bf p}$, so that ${\bf k} = f({\bf p})$ transforms under the normal Lorentz-transformation~$\Lambda$.
In the literature, ${\bf k}$ is often referred to as the `pseudo-momentum'~\cite{Judes:2002bw}.
Here and in the following bold-faced quantities denote vectors, and so the function~$f$ is a~map from ${\mathbb{R}}^4
\to {\mathbb{R}}^4$.
It can be shown that the opposite is also true: given the function~$f$ it is possible to construct the non-linear
Lorentz-transformation $\widetilde \Lambda$, so these two formulations of {\sc DSR} are actually equivalent to each
other~\cite{Hossenfelder:2005ed}.

\looseness=-1
It is then straightforward to construct the modif\/ied addition law by use of the pseudo-momentum $\mathbf{k}$ that by
assumption transforms under the normal Lorentz transformation.
To each momentum ${\bf p}$ we have an associated pseudo-momentum $\mathbf{k}_1 = f(\mathbf{p}_1)$, $\mathbf{k}_2=f(\mathbf{p}_2)$.
The sum $\mathbf{k}_1 + \mathbf{k}_2$ is invariant under normal Lorentz transformations, and so we construct the sum of
the $\mathbf{p}$'s as
\begin{gather*}
\mathbf{p}_1 \oplus \mathbf{p}_2 = f^{-1}(\mathbf{k}_1 + \mathbf{k}_2) = f^{-1}(f(\mathbf{p}_1) + f(\mathbf{p}_2)).
\end{gather*}
This is one way to arrive at the modif\/ied addition law for momenta.
It requires one to f\/irst construct the pseudo-momentum.
In some cases it is easier to extract the modif\/ied addition law from an algebraic approach that starts with the modif\/ied
commutation relations in the Poincar\'e-algebra; it is the coproduct of the~$\kappa$-Poincar\'e
algebra~\cite{KowalskiGlikman:2003we}.

The new sum is in general not associative.
If one does not symmetrize the new addition rule, the result of the new sum may also depend on the order in which
momenta are added, ie $({\bf p}_1 \oplus {\bf p}_2) \oplus {\bf p}_3 \neq {\bf p}_1 \oplus ({\bf p}_2 \oplus {\bf p}_3)$.
This means in particular the sum of two momenta can depend on a~third term that may describe a~completely unrelated (and
arbitrarily far away) part of the universe, which has been dubbed the `spectator
problem'~\cite{Girelli:2004ue, KowalskiGlikman:2004qa}.
We will not further address this problem here.

\looseness=-1
This new def\/inition for a~sum is now nicely observer-independent by construction, but brings with it a~new problem.
The non-linear contributions in~$f$ by construction~-- going back to the motivation for {\sc DSR}~-- become relevant
when the momenta (or some of its components respectively) come close by the Planck energy.
For a~single elementary particle these nonlinear contributions will be small.
But the (usual) total momentum of a~collection of particles can easily exceed the Planck energy.
The Planck mass is a~large energy as far as particle physics is concerned, but in everyday units it is about $10^{-5}$
gram, a~value that is easily exceeded by some large molecules.
This problem of reproducing a~sensible multi-particle limit when one chooses the physical momentum to transform under
modif\/ied Lorentz transformations has become known as the `soccer-ball problem', where the soccer-ball stands in for any
object exceeding the Planck mass.

The soccer-ball problem can be sharpened as follows.
If one makes an expansion of the function~$f$ to include the f\/irst correction terms in ${\bf p}/m_{\mathrm{Pl}}$, and
from that derives the sum $\oplus$, then it remains to be shown that the correction terms stay smaller than the linear
terms if one calculates sums over a~large number of momenta.
One expects that the sum then has approximately the form $\mathbf{p}_1 \oplus \mathbf{p}_2 \approx \mathbf{p}_1 +
\mathbf{p}_2 + \mathbf{p}_1 \mathbf{p}_2 \Gamma/m_{\mathrm{Pl}}$, where~$\Gamma$ are some coef\/f\/icients of order one.
If one iterates this sum for~$N$ terms, the normal sum grows with~$N$ but the number of additional terms scales with
${\sim}N^2$ for $N\gg1$.
To solve the soccer-ball problem, the non-linear contributions have to grow slower than~$N$ in spite of this, and are
still small compared to the linear term even when the total momentum exceeds the Planck scale.

One may slightly weaken this requirement because the universe does not contain an inf\/inite amount of particles, so
strictly speaking the non-linear terms only have to remain negligible for collections of particles that we have
observed.
However, we note that once one ventures into the realm of quantum f\/ield theory~-- which is necessary eventually to
reproduce the physics we know~-- the number of constituent particles becomes ill-def\/ined.
It can plausibly be said that the proton does in fact contain an inf\/inite amount of particles, and is composed of
a~whole sea of virtual quarks rather than just three valence quarks.
If one only solves the soccer-ball problem for f\/inite~$N$, it is thus bound to come back once one attempts to treat
quantized interactions.

\section{Proposed solutions}

There have been various attempts to address the soccer-ball problem, but so far none has been generally accepted.

For example, it has been suggested that with the addition of~$N$ particles, the Planck scale that appears in the
non-linear Lorentz transformation and in the modif\/ied addition law should be rescaled
to $m_{\mathrm{Pl}}N$~\cite{Judes:2002bw,Magueijo:2006qd, Magueijo:2002am}.
It is however dif\/f\/icult to see how this ad-hoc solution could follow from the theory.
In~\cite{Liberati:2004ju} it has been proposed to reinterpret {\sc DSR} as a~modif\/ied theory of measurement, thereby
also addressing the soccer-ball problem.

Alternatively, it has been proposed that the scaling of modif\/ications should go with the
density~\cite{Hossenfelder:2007fy,Olmo:2011sw} rather than with the total momentum or energy respectively.
While the energy of macroscopic objects is larger than that of microscopic ones, the energy density decreases instead.
This seems a~natural solution to the issue but would necessitate a~completely dif\/ferent ansatz to implement.

The so far most promising approach has been put forward in~\cite{AmelinoCamelia:2011uk} in the context of relative
locality.
It was claimed in this paper that the soccer ball problem is absent for a~particular choice of connection $\Gamma_\mu^{\;
\alpha \beta}$ and basis on the manifold that is momentum space.
This basis are normal coordinates in which the connection is totally antisymmetric $\Gamma_\mu^{\;
(\alpha \beta)} = 0$ with the consequence that collinear momenta add under the normal addition law:
\begin{gather*}
\mathbf{p}_1 \oplus \mathbf{p}_2 = \mathbf{p}_1 + \mathbf{p}_2
\qquad
\text{if}
\qquad
{\bf p}_1 \parallel {\bf p}_2.
\end{gather*}

Now in the f\/irst conceptions of relative locality, the connection~$\Gamma$ was a~metric connection on the momentum space
manifold.
In that case the requirement of the existence of global normal coordinates means that the curvature of momentum space
actually has to vanish and only the torsion is non-trivial.
This is in stark contrast to the momentum space geometries that have previously been considered, typically Anti-de
Sitter spaces with vanishing torsion.
Alternatively, the connection~$\Gamma$ is not a~metric connection but a~connection deriving from the structure of the
group and its addition law.
This interpretation is natural when dealing with a~group manifold.
Alas, when dealing with a~general manifold like one does in relative locality it is not a~priori clear that normal
coordinates exist globally.

However, having mentioned this caveat we will in the following assume the coordinates exist and interpret~$\Gamma$ as
a~general, not necessarily a~Riemannian, connection on the manifold.
Note however, that in the relative locality approach the Lagrangian is constructed from the geodesic distance of momenta
from the origin, and for that one needs the metric connection.

Using the normal coordinate basis then, it was shown in~\cite{AmelinoCamelia:2011uk} that the momentum exchange between
two macroscopic bodies (`soccer-balls') does not pose a~problem provided the total momentum of the bodies themselves is
well-def\/ined.
However, it was argued in~\cite{Hossenfelder:2012vk} that the total momentum of the macroscopic bodies is in fact not
well-def\/ined: Assuming a~body of a~certain temperature~$T$, the f\/luctuations in the constituents' momenta are not
aligned and random in direction.
The f\/luctuations do thus contribute to the modif\/ied sum in the normal coordinates even though the average momenta don't.
It was estimated in~\cite{Hossenfelder:2012vk} on fairly general grounds that the non-linear terms scale with $N^{3/2}$
and thus the sum does not converge.

In the reply~\cite{Amelino-Camelia:2013zja} if was pointed out that this estimate neglected some second-order terms that
mix into the f\/irst order-terms.
The more detailed analysis leads to a~constraint on the connection coef\/f\/icients that has to be fulf\/illed in order for
the terms that scale with powers larger than~$N$ to be absent.
This analysis is a~big step forward.
However, it remains to be shown that there is any non-trivial manifold that fulf\/ils these requirements and that not
taking into account subsequently higher orders will leave only the linear addition law as option.
It must be kept in mind that by the construction of {\sc DSR} one expects all the additional terms in the non-linear sum
to asymptotically scale with~$N$ and not any faster, so that they do exactly cancel the growth of the linear term.
This makes it quite implausible that there is a~maximum power of~$N$ in the expansion.
Alas, as previously mentioned, as long as one does not deal with quantized interaction, taking $N \to \infty$ might not
be required.

That having been said, the approach to the soccer-ball problem pursued in~\cite{Amelino-Camelia:2013zja} is technically
rigorous and of\/fers the possibility of settling the issue at least for the normal coordinates.
What is left to show is that the coordinate system exists globally and that the requirements for absence of the
divergent terms can be fulf\/illed by any non-trivial addition laws.

\section{Composite systems and statistical mechanics}

Another way to deal with the soccer-ball problem that has become very popular is to just ignore it.
Many approaches to the description of composite systems or many particle states have been made based on the modif\/ied
commutation relations either without subscribing to the deformed Lorentz transformations, and thereby generically
breaking Lorentz invariance, or by employing an ad hoc solution by rescaling the bound on the energy with the number of
constituents, or by not rescaling the bound and thus f\/inding -- not so surprisingly -- that Planck scale ef\/fects become
noticeable in composite objects with masses approaching the Planck mass.

For example, it was proposed recently that a~massive quantum mechanical oscillator might allow one to test Planck-scale
physics~\cite{Pikovski:2011zk} in a~parameter range close to the Planck scale.
This conclusion was reached not taking into account that, to circumvent the soccer-ball problem, the Planck-mass has to
be rescaled with the number of constituents.
If one does not do this rescaling the model does not even reproduce classical Newtonian mechanics because the momenta of
macroscopic objects do no longer add linearly, and the violations of linearity are dominant.
The experiment proposed in~\cite{Pikovski:2011zk} thus tests a~model we already know is wrong.
This is also so if one does not subscribe to the deformed Lorentz-transformation and instead considers the modif\/ied
addition law to just violate observer-independence because Lorentz-invariance violation is extremely strongly
constrained already~\cite{Kostelecky:2008ts}, far beyond the Planck scale.

That having been said, one can of course investigate the statistical mechanics and thermodynamics on a~purely
mathematical footing and just leave the deformation scale unspecif\/ied, so that its relation to the Planck scale and the
number of constituents can be taken into account later, once the soccer-ball problem has been solved.

In that spirit, the statistical mechanics from the~$\kappa$-Poincar\'e algebra was investigated in general
in~\cite{Fityo:2008zz, KowalskiGlikman:2001ct}.
The partition functions of deformed quantized statistical mechanics have been derived in~\cite{Pedram:2011aa},
in~\cite{Ali:2011ap} the consequences for the Liouville theorem were investigated, and in~\cite{Chang:2001bm} the
modif\/ication of the density of states and the arising consequences for black-hole thermodynamics were studied.
In~\cite{Quesne:2009vc} corrections to the ef\/fective Hamiltonian of macroscopic bodies have been considered and
in~\cite{KalyanaRama:2001xd} statistical mechanics with a~generalized uncertainty and possible applications for
cosmology have been looked at.~\cite{Nozari:2006gg} studied the thermodynamics of ultra-relativistic particles in the
early universe and relativistic thermodynamics has been investigated in~\cite{Das:2009qb}.~\cite{Wang:2011iv} studied
the equation of state for ultra-relativistic Fermi gases in compact stars, the ideal gas was addressed
in~\cite{Chandra:2011nj} and photon gas thermodynamics in~\cite{Zhang:2011ms}.

As previously mentioned, the physical consequences of these investigations should be interpreted with caution whenever
energies have not been rescaled with the number of constituents.

\section{Summary}

The idea that Lorentz-symmetry might be modif\/ied so that the Planck energy becomes observer-independent is interesting
and leads to testable phenomenological consequences.
But consistently modifying Special Relativity is hard.
We have seen here that such a~deformation of Special Relativity necessitates a~non-linear addition law for momenta and
it is presently not known whether a~sum of a~large number of momenta converges suitably.
If it does not converge, these models cannot reproduce even classical mechanics.
This so-called `soccer-ball problem' is thus a~pressing one,
for without solving it models with deformed Lorentz-symmetry fail to reproduce well-tested physics.

A recent approach within the scenario of relative locality seems promising.
It has been shown that there exist addition rules which do not suf\/fer from the problem and concrete requirements for the
absence of the soccer-ball problem have been derived in this context.
It yet remains to be seen however whether in the class of these addition rules which are immune to the soccer-ball
problem are any which are in fact non-linear and do not just reproduce Special Relativity.

Finally, we note that the soccer-ball problem does not occur in the case in which the modif\/ication of the
Lorentz-transformation only occurs for of\/f-shell momenta which seems to be suggested in some
interpretations~\cite{Hossenfelder:2006cw}.
Then, if one identif\/ies the momenta of particles as those of the asymptotically free states, the addition of their
momenta is linear as usual.
For the same reason, the problem also does not appear in the interpretation of such modif\/ications of conservation laws
as being caused by a~running Planck's constant, put forward~\cite{Calmet:2010tx, Percacci:2010af}, and discussed
in~\cite{Hossenfelder:2012jw}.
In this case, the relevant energy is the momentum transfer, and for bound states this remains small even if the total
mass increases beyond the Planck mass.

\subsection*{Acknowledgements}

I thank Jerzy Kowalski-Glikman for helpful conversation.

\pdfbookmark[1]{References}{ref}
\LastPageEnding

\end{document}